\pdfoutput=1

\documentclass[10pt,twocolumn]{article}
\usepackage{neutral}

\makeatletter
\renewcommand{\@maketitle}{%
  \vbox{%
    \hsize\textwidth
    \linewidth\hsize
    \centering
    \vskip 0.1in
    {\Large\bfseries \@title \par}
    \vskip 0.2in
    \def\And{%
      \end{tabular}\hfil\linebreak[0]\hfil%
      \begin{tabular}[t]{c}\bfseries\rule{\z@}{24\p@}\ignorespaces%
    }%
    \def\AND{%
      \end{tabular}\hfil\linebreak[4]\hfil%
      \begin{tabular}[t]{c}\bfseries\rule{\z@}{24\p@}\ignorespaces%
    }%
    {\large\begin{tabular}[t]{c}\bfseries\rule{\z@}{24\p@}\@author\end{tabular}}%
    \vskip 0.6in \@minus 0.1in
  }%
}
\makeatother

\renewenvironment{abstract}
  {\centerline{\large\bfseries\abstractname}\begin{quote}}
  {\end{quote}}

\AtBeginDocument{%
  \newgeometry{textheight=9in,textwidth=520pt,top=1in,
               headheight=14pt,headsep=25pt,footskip=30pt}%
  \setlength{\columnsep}{16pt}}

\usepackage[T1]{fontenc}
\usepackage[utf8]{inputenc}
\usepackage{microtype}

\usepackage[round]{natbib}
\usepackage{graphicx}
\usepackage{amssymb}
\usepackage[fleqn]{amsmath}
\usepackage{amsthm}
\usepackage{breqn}
\usepackage{comment}
\usepackage{tikz}
\usepackage{pgfplots}
\usetikzlibrary{external,positioning,shapes,arrows, calc, fit, matrix, arrows.meta, decorations.pathreplacing, shapes.multipart, calligraphy}
\usepackage{float}
\usepackage{booktabs}
\pgfplotsset{compat=1.18}

\usepackage{import}
\usepackage{catchfile}

\newcounter{formdef}

\usepackage[breaklinks]{hyperref}
\definecolor{darkblue}{rgb}{0, 0, 0.5}
\hypersetup{colorlinks=true, citecolor=darkblue, linkcolor=darkblue,
            urlcolor=darkblue}
\usepackage[nameinlink]{cleveref}
\crefname{formdef}{Problem}{Problems}
\Crefname{formdef}{Problem}{Problems}
\crefname{section}{Section}{Sections}
\crefname{subsection}{Section}{Sections}
\crefname{table}{Table}{Tables}
\crefname{equation}{Equation}{Equations}
\creflabelformat{equation}{#2#1#3}

\makeatletter
\AtBeginDocument{%
  \let\breqn@orig@setnumber\eq@setnumber
  \def\eq@setnumber{%
    \let\breqn@orig@stepcounter\stepcounter
    \def\stepcounter##1{%
      \let\stepcounter\breqn@orig@stepcounter \refstepcounter{##1}}%
    \breqn@orig@setnumber
    \let\stepcounter\breqn@orig@stepcounter}%
}
\makeatother

\title{Interpretable Memory Models for Spaced Repetition}

\makeatletter
\newcounter{paperauthor}
\newcommand{\paperauthors}{}
\newif\if@paperauthorfirst
\newcommand{\authorline}[1]{%
  \def\@paperauthorline{#1}%
  \ifx\@paperauthorline\@empty\else
    \if@paperauthorfirst
      \@paperauthorfirstfalse
    \else
      \g@addto@macro\paperauthors{\\}%
    \fi
    \g@addto@macro\paperauthors{#1}%
  \fi}

\newcommand{\paperauthornames}{}
\newcommand{\authorname}[1]{%
  \ifnum\value{paperauthor}>1 \g@addto@macro\paperauthornames{, }\fi
  \g@addto@macro\paperauthornames{#1}%
  \authorline{#1}}
\newcommand{\authoremail}[1]{\authorline{\texttt{#1}}}
\newcommand{\paperauthor}[1]{%
  \stepcounter{paperauthor}%
  \ifnum\value{paperauthor}>1 \g@addto@macro\paperauthors{\And}\fi
  \@paperauthorfirsttrue
  #1}
\paperauthor{%
  \authorname{Anders Schill}%
  \authorline{Stockholm, Sweden}%
  \authoremail{andersjschill@gmail.com}%
}

\makeatother

\author{\paperauthors}

\hypersetup{pdftitle={Interpretable Memory Models for Spaced Repetition},
            pdfauthor={\paperauthornames},
            pdfkeywords={spaced repetition, memory models}}

\newcommand{\fittocolumn}[1]{\resizebox{\columnwidth}{!}{#1}}

\makeatletter

\newcommand{\paperappendices}[1]{%
  \appendix
  \renewcommand{\@seccntformat}[1]{\appendixname~\thesection:\ }%
  \def\paperappendix##1##2{\section{##1}\import{./content/}{##2}}%
  #1}
\makeatother

\newcommand{\paperfrontmatter}[1]{%
  \twocolumn[\maketitle]%
  \begin{abstract}#1\end{abstract}}

\begin{document}

\paperfrontmatter{Spaced repetition software schedules reviews with a memory model fit to review logs.
Accuracy on a test set is not sufficient evidence of quality since
available data are produced by existing schedulers, and new solutions must extrapolate beyond them.
A model also needs a simple mechanistic interpretation.
We present SBD, a model that is more interpretable and 
80\% smaller than the current state of the art 
at nearly the same accuracy.
}

\section{Introduction}
Spaced repetition is an important component in many types of educational software,
but perhaps becomes most intuitively clear in the context of digital flashcards.
In these applications, a student is trying to memorize a large deck of cards,
each of which contains a small piece of information such as a vocabulary term.
A spaced repetition algorithm decides when to introduce new cards and when to review old ones
with the goal of helping the student memorize the deck efficiently.
Every time the student sees a card, they rate how well they remembered it, and
the software maintains a log with these ratings similar to \cref{review-log-table}.

\begin{table}[H]
\centering
\begin{tabular*}{\columnwidth}{l@{\extracolsep{\fill}}rrr}
\hline
\textbf{Timestamp} & \textbf{Card ID} & \textbf{Rating} & \textbf{Response Time} \\
\hline
\text{1786231271076} & 89392 & 2 & 8369 \\
\text{1786231285712} & 23232 & 4 & 2250 \\
\text{1786231291192} & 96654 & 4 & 5654 \\
\text{1786231299176} & 26813 & 1 & 6664 \\
\text{1786231315483} & 10557 & 3 & 6860 \\
\hline
\end{tabular*}
\caption{Example of a review log}
\label{review-log-table}
\end{table}

Modern spaced repetition algorithms have two parts,
a memory model that predicts the probability that a student still remembers 
a card at some specified future time, and a scheduler
that uses the memory model to decide what cards to show when.

Available data only cover the intervals that existing schedulers chose,
while a new scheduler must search outside them.
As discussed in \cref{sec:model-constraints}, accuracy on a test set therefore
says little on its own, and a model also needs a mechanistic interpretation to
be trusted beyond the data.
In this paper, we present SBD, a model with near state-of-the-art performance
despite being substantially smaller and more interpretable.
We also show that common formulations of the scheduling problem have simple, 
near-optimal solutions given this model.%
\footnote{Code is available at \url{https://github.com/andersschill/memory-model-benchmark-for-spaced-repetition}.}

\section{Related Work}
The spaced repetition literature is split between psychology,
machine learning, and practical implementations that were never formally
published.

\subsection{Early Systems}
\citeauthor{ebbinghaus1885} measured how memory decays over time as early as
\citeyear{ebbinghaus1885}, but the first learning systems were mostly pure heuristics
that output intervals directly without any optimization or memory modeling.
\citet{pimsleur1967} prescribes fixed, exponentially spaced intervals
for reviews, and \citet{leitner1972} adds that failure should send cards back to the
beginning. \citet{atkinson1972} is the exception. They solve a different problem, ordering
items within a single study session rather than deciding when to review them
later, but the approach is modern in the sense that it has a memory model fitted
to data and a scheduler that plans over its state against an explicit objective.

\subsection{SuperMemo}
SuperMemo, released in 1987, was an early software application for digital flashcards.
It shipped with a scheduling algorithm known as SM-2, where every card carries 
an interval $I$ and an easiness factor $d$ \citep{sm2_1990}.
Each review multiplies the interval and moves the
easiness by a fixed, hand-picked function $f$ of the rating $g$.
\begin{equation*}
I' = d\,I, \qquad d' = d + f(g) .
\end{equation*}
The output is the interval itself, so the algorithm schedules without ever
predicting recall. 
\citet{wozniak1994} replaces the single multiplier $d$ with a table
of multipliers indexed by repetition number and easiness, each entry
adjusted during use until the retention measured at that entry reaches a target.
\citet{wozniak1995} makes the design explicit by separating scheduling from the memory model
and proposing a system that recovers the SM-2 intervals when cards are scheduled at a constant retention.
\begin{equation*}
s' = d\,s, \qquad R(t) = e^{-t/s} .
\end{equation*}
\citeauthor{sm2_1990} has kept updating the algorithm, with the latest version being SM-20,
but keeps the full definitions proprietary.
From wiki pages with partial descriptions \citep{sm17_wiki, sm18_wiki, sm20_wiki}, 
we learn that the models still have the same two state variables,
that SM-20 has around 40 weights, and that the models before it
were based on tabular transitions rather than given a parametric form.

\subsection{Models without Schedulers}
A large literature aims mainly at making the memory model more accurate, and
treats scheduling, if at all, as a secondary concern.
These papers predict recall from the review history and
differ mainly in the functional form. ACT-R sums a power law decay over
every past review \citep{pavlik2005} and MCM runs a cascade of leaky integrators
\citep{mozer2009}, both fit to laboratory experiments rather than logs. DASH and
DAS3H are logistic regressions on practice counts in trailing windows
\citep{lindsey2014, choffin2019}, \citet{hlr2016} fit the exponential
that \citet{wozniak1995} assumed to actual review logs,
and neural networks push the same regression further \citep[e.g.,][]{zaidi2020,
ye2022lstm, pokrywka2023, imem2026}, sometimes extracting
a closed form equation back out for interpretability
\citep[e.g.,][]{psyinn2024, spsyinn2025, rokan2025}.
Ebisu \citep{ebisu} fits nothing at all, and instead updates a
per-card Beta posterior over recall probability by moment matching.

\subsection{DHP and FSRS}
\label{sec:related-fsrs}

\citet{ye2022} and its journal extension \citet{ye2023} build on
\citet{wozniak1995} to create a full practical system.
Their DHP model keeps
the same two state variables, a half-life and a difficulty, with update rules
whose weights are fit to review logs. Their scheduler
minimizes the reviews needed to reach a target stability, with dynamic
programming over a discretized state space.

The work by \citeauthor{ye2022} led to the Open Spaced Repetition project
\citep{open_spaced_repetition} and its algorithm FSRS.
FSRS is both a memory model and a scheduler based on the paper,
but the memory model has evolved.
The current version, FSRS-7 \citep{srs_benchmark}, adds a short-term stability as a third state
variable and has 34 weights.
Its forgetting curve is a mixture of two Lomax survival functions
\citep{lomax1954}, one per stability.

\subsection{Benchmark and Dataset}
The Open Spaced Repetition project has released a public dataset of
10,000 review logs \citep{anki_revlogs_10k},
and their repository \citep{srs_benchmark} has become a public leaderboard
for memory models.

FSRS-7 currently defines the accuracy-parsimony frontier of
\cref{sec:model-constraints}, and is the closest in nature to the models we
present in this paper.

\section{Scheduling}
\subsection{Problem Definition}
\label{sec:problem-definition}

Formal definitions of the scheduling problem vary, but at their core
there is a tradeoff between learning and effort.
In \cref{sec:planning}, we maximize the area under the deck's recall curve, discounted over time,
and represent the effort constraint as either a daily budget or a cost per review.

\subsection{Planning vs Reinforcement Learning}
\label{sec:scheduling-approaches}

Scheduling flashcards is a sequential decision problem,
and besides simple heuristic methods, there are two main approaches to solving it in the literature:
reinforcement learning \citep[e.g.,][]{reddy2017, upadhyay2018, xiao2024, zhai2025} and
planning \citep[e.g.,][]{reddy2016, tabibian2019, ye2022, open_spaced_repetition}.
Both need a memory model, but they use it in different ways.
Reinforcement learning learns a policy by simulating reviews,
and planning optimizes over the model's state directly.
The latter imposes additional constraints on the model, as we discuss in \cref{sec:model-constraints}.

\subsection{Constraints on the Memory Model}
\label{sec:model-constraints}

Like in many other forecasting domains, large sequential neural networks or tree
ensembles with clever feature engineering are the most accurate.
In practice, however, the following constraints disqualify them.

\paragraph{Reliable extrapolation.}
Available training data come from educational software where a scheduler
decided the review intervals. Consequently, the data only cover a narrow band.
A new scheduler searches the entire range, and the model has to extrapolate.
A test set comes from the same band, and therefore cannot tell which model does so
reliably. This problem is classic. The Lucas critique says that a model fit
under one policy need not hold under another \citep{lucas1976}.
Accuracy is therefore necessary but not sufficient.
The solution is parsimony, a model with no more structure than the
mechanism it describes, where every term has a reason to be there beyond the data.

\paragraph{Simple state dynamics.}
Planning optimizes over the model's state, either by discretizing it or by
working with it analytically.
A discretized state space grows exponentially in the number of state variables,
and analytic methods require transitions built from a few elementary operations.
Either way, a model with a large or opaque state cannot be planned over.

\paragraph{Self-contained state updates.}
To plan over hypothetical future reviews, a planner or simulator has to roll the model forward.
Every input the model consumes must be something the model itself produces.
This, for example, rules out any feature measured after the last review.

These constraints leave a fairly narrow class of models, and within it,
there is still a tradeoff between accuracy and parsimony. 
Advancing that tradeoff's frontier is the problem we address in this paper.

\section{The SBD Model}
We call our model SBD, after its three state variables: stability $s$,
brittleness $b$, and difficulty $d$, which evolve according to the
equations below.
In these equations, $t$ is the number of days elapsed since the previous
review, $q \in \{1,\dots,4\}$ the rating, $q_0$ the rating of the
first review, and $w_1$--$w_7$ the fitted weights. Inside
\cref{eq:stability}, $R$ is the recall \eqref{eq:curve} at the moment
of review.
\begingroup
\setlength{\jot}{8pt}
\setlength{\abovedisplayskip}{14pt}
\setlength{\belowdisplayskip}{8pt}
\begin{flalign}
& a = t^{1/w_1} + \sqrt{b\,s} && \label{eq:age}
\\
& \alpha = w_2\, d^{\,w_3} && \label{eq:decay}
\\
& R\!\left(t\right) = \left(1 + \frac{a}{s}\right)^{-\alpha} && \label{eq:curve}
\\
& s' = s \left\{
\begin{array}{@{}l@{\;}l@{}}
  R^{-1/\alpha} & \scriptstyle q \ge 2\\[3pt]
  \left(1-R\right)^{w_3} & \scriptstyle q = 1
\end{array}
\right. && \label{eq:stability}
\\
& b' = w_4\, w_5^{\,1-q} && \label{eq:brittleness}
\\
& d' = d + \max\!\left(3-q,\,0\right) && \label{eq:difficulty}
\\
& \left(s_0, b_0, d_0\right) = \nonumber \\
& \qquad \left( w_6\, w_5^{\,q_0-1},\; w_4\, w_5^{\,1-q_0},\; w_7 + \max\!\left(3-q_0,\,0\right) \right) && \label{eq:init}
\end{flalign}
\endgroup

\paragraph{\protect\Cref*{eq:age} (age).} The age $a$ is the time
since the last review. It is compressed by the exponent $1/w_1$ so that the
card is forgotten more slowly the longer it survives, consistent with Jost's law
\citep{jost1897}. A review is modeled as an imperfect repair via $\sqrt{b\,s}$, 
which is the concept of a virtual age from \citet{kijima1989}.

\paragraph{\protect\Cref*{eq:decay} (decay).} The decay $\alpha$ is
the exponent of the forgetting curve. It grows with difficulty, so a
difficult card is forgotten faster at every age.

\paragraph{\protect\Cref*{eq:curve} (forgetting curve).}
We model the rate at which a card is forgotten as Gamma distributed
with shape $\alpha$ and scale $1/s$. The chance that the card has
survived to age $a$ is then \cref{eq:curve}. This is Gamma
frailty \citep{vaupel1979}. Without the repair term $\sqrt{b\,s}$,
\cref{eq:curve} is the survival function of a Burr XII distribution
in $t$ \citep{burr1942}, and Wickelgren's power law of forgetting
\citep{wickelgren1974}, with its interference factor dropped, is then
the special case $w_1 = 1$.

\paragraph{\protect\Cref*{eq:stability} (stability update).} A pass
is the observation that the card survived to age $a$. The Bayesian
update on that observation keeps $\alpha$ and replaces $s$ with
$s + a$, which is the pass row.
A lapse is damage rather than evidence. Stability shrinks by the
surprise of the failure. An expected failure erases little and a
surprising one erases more.

\paragraph{\protect\Cref*{eq:brittleness} (brittleness update).}
Brittleness describes how incomplete the repair of the last review was.
It depends only on the rating and is highest after a failure and lowest after an easy
pass.

\paragraph{\protect\Cref*{eq:difficulty} (difficulty update).}
Difficulty is a tally of a card's failures and hard passes.

\subsection{The Binary SBD Model}
\label{sec:stim-binary}

We obtain a binary version of the SBD model by training the full model on
binary ratings. It still uses 7 parameters, but its state updates are self-contained,
as discussed in \cref{sec:model-constraints}.

\subsection{Fine-Tuning}
\label{sec:finetuning}

A few non-standard fine-tuning strategies have been shown to improve FSRS on
the benchmark \citep{srs_benchmark}, and we adopt two of them. Recent reviews are weighted
more heavily, and a Gaussian prior pulls each weight toward its pretrained
value. We set the width of that prior from the spread of unregularized per-user
fits, and let its strength fall as the user accumulates reviews.

\section{Results}
\label{sec:results}
\subsection{Data and Evaluation}
\label{sec:data}

We use the anki-revlogs-10k dataset \citep{anki_revlogs_10k},
which consists of 10,000 review logs collected by 
the Open Spaced Repetition project.
Each review carries a rating from 1 to 4.
A rating of 1 means the card was forgotten and a rating of 2 or higher
means it was recalled.

Users are split into training (50\%), validation (25\%), and test (25\%) sets.
A model is pretrained once on the training set, and then evaluated on held-out
users.
For fine-tuning, we follow the protocol of the Open Spaced Repetition project's public benchmark
\citep{srs_benchmark}. Every user's reviews are split in time into five
blocks, and each block is predicted by a model trained on the reviews before it.

\subsection{Prediction Accuracy}
\begin{table}[H]
\centering
\small
\setlength{\tabcolsep}{4pt}
\fittocolumn{%
\begin{tabular}{lcrrrrr}
\hline

\textbf{Model} & \multicolumn{1}{l}{\shortstack[l]{\textbf{Fine-}\\\textbf{Tuned}}} & \multicolumn{1}{l}{\textbf{Params}} & \multicolumn{1}{l}{\shortstack[l]{\textbf{State}\\\textbf{Dims}}} & \multicolumn{1}{l}{\textbf{BCE}} & \multicolumn{1}{l}{\shortstack[l]{\textbf{RMSE}\\\textbf{(bins)}}} & \multicolumn{1}{l}{\textbf{AUC}} \\

\hline

\text{FSRS-7} & x & 34 & 3 & 0.3237 & 0.0577 & 0.7517 \\
\text{SBD} & x & 7 & 3 & 0.3279 & 0.0593 & 0.7441 \\
\text{SBD Binary} & x & 7 & 3 & 0.3296 & 0.0598 & 0.7329 \\
\text{FSRS-7} &  & 34 & 3 & 0.3454 & 0.0900 & 0.7282 \\
\text{SBD} &  & 7 & 3 & 0.3469 & 0.0878 & 0.7224 \\
\text{SBD Binary} &  & 7 & 3 & 0.3483 & 0.0887 & 0.7057 \\

\hline
\end{tabular}%
}
\caption{Results on the test set}
\label{results-table}
\end{table}

\Cref{results-table} compares SBD to FSRS-7 (\cref{sec:related-fsrs}) on BCE,
RMSE(bins), and AUC on the test set,
computed per user and averaged over users rather than pooled.
RMSE(bins) is a calibration metric developed by the benchmark
\citep{srs_benchmark}.
The benchmark reports results both with and without same-day repeated
reviews of a card, and we use the variant that
includes them.

Fine-tuned, SBD is around 1\% behind FSRS-7 on BCE and AUC, and 3\% on RMSE(bins).
With only pretraining, the differences in BCE and AUC are even smaller, and SBD outperforms FSRS-7 on RMSE(bins) by 2\%.

By the argument in \cref{sec:model-constraints}, model complexity matters
as much as accuracy, and here SBD has the edge.
It has 80\% fewer weights than FSRS-7, and arguably a clearer mechanistic interpretation.

\section{Planning with SBD}
\label{sec:planning}
In this section, we describe planning solutions to the scheduling 
problem defined in \cref{sec:problem-definition}.
We plan with binary SBD rather than the full model. 
Its state updates are self-contained, as discussed in \cref{sec:model-constraints}, 
whereas the full model takes ratings as input without predicting them. 
Planning with the full model would need a separate distribution over ratings.

The objective is to maximize the area under the deck's recall curve, discounted over time.
Write $x = \left(s, b, d\right)$ for the state of a card, and let
$x_{\mathrm{pass}}\!\left(t\right)$ and $x_{\mathrm{fail}}\!\left(t\right)$
be the states that binary SBD gives after a pass or a fail at interval $t$.
With $\lambda$ denoting the cost of a review and $\gamma$ the daily discount factor,
the value of a card is given by the infinite-horizon discounted Bellman equation
\begin{equation*}
\begin{aligned}
V\!\left(x\right) &= \max_{t}\;
\int_0^t \gamma^{\tau} R\!\left(\tau\right) \mathrm{d}\tau \\
&\; + \gamma^{t} \Bigl[ R\!\left(t\right) V\!\left(x_{\mathrm{pass}}\!\left(t\right)\right)
+ \left(1 - R\!\left(t\right)\right) V\!\left(x_{\mathrm{fail}}\!\left(t\right)\right)
- \lambda \Bigr],
\end{aligned}
\end{equation*}
which is trivial to solve with value iteration.
The maximizing interval, $t^{*}_{\lambda}\!\left(x\right)$, is
the optimal interval at review cost $\lambda$.

If we instead have a budget of $B$ reviews per day, cards compete for slots,
and the problem becomes a restless bandit.
This problem is PSPACE-hard in general \citep{papadimitriou1999},
but the standard approach is the Whittle relaxation
\citep{whittle1988, weber1990index}.
It assigns each card an index $\eta$, the review cost at which the
student is indifferent between reviewing the card today and waiting.
Since the optimal interval grows with the review cost, a card in state $x$
that has waited $t$ days has a unique index, namely the cost $\eta$ at
which $t$ is the optimal interval,
\begin{equation*}
t^{*}_{\eta}\!\left(x\right) = t.
\end{equation*}
The policy that follows is to review each day the $B$ cards with the
largest index, which naturally also handles the introduction of new cards.
As the deck grows with the budget kept proportional, index policies of
this kind are asymptotically optimal under an indexability and stability
condition \citep{weber1990index}.

\section{Conclusion}
As argued in \cref{sec:model-constraints}, test set performance alone cannot 
separate models in this space. SBD is considerably more parsimonious, 
at almost no cost to performance compared to FSRS-7.
It is certainly possible to modify the SBD model to gain more accuracy, 
but any such gain must be weighed against a loss in interpretability.

Running live A/B tests that compare SBD to FSRS-7 and planning to reinforcement
learning would be a natural next step, since simulation offers no clean way to
compare solutions end-to-end. Simulators are themselves built on memory models, and beyond that,
they cannot capture how user behavior shifts in a new environment. 
However, for practical reasons, A/B tests
between state-of-the-art algorithms are absent from the spaced repetition
literature in general.

\bibliography{references}

\paperappendices{%
  \paperappendix{Fitted Weights}{9_appendix.tex}%
}

\end{document}